\documentclass[twocolumn]{aastex631}

\usepackage{amsmath}
\usepackage{bm}

\usepackage{makecell}

\shorttitle{HI-rich `Dark' Galaxies in \textsc{hestia} \& Auriga}
\shortauthors{Zheng et al.}

\begin{document}

\defcitealias{Benitez-Llambay2017}{BL17}

\title{HIDES - I. The population and diversity of HI-rich `dark' galaxies in the Hestia and Auriga simulations} 

\author[0000-0002-1665-5138]{Haonan Zheng}
\affiliation{Kavli Institute for Astronomy and Astrophysics, Peking University, Beijing 100871, China}

\author[0000-0001-6115-0633]{Fangzhou Jiang}
\affiliation{Kavli Institute for Astronomy and Astrophysics, Peking University, Beijing 100871, China}

\author[0000-0001-7075-6098]{Shihong Liao}
\affiliation{Key Laboratory for Computational Astrophysics, National Astronomical Observatories, Chinese Academy of Sciences, Beijing 100101, China}

\author[0000-0002-6406-0016]{Noam I. Libeskind}
\affiliation{Leibniz-Institut für Astrophysik Potsdam (AIP), An der Sternwarte 16, 14482 Potsdam, Germany}

\correspondingauthor{Haonan Zheng} \email{hnzheng@pku.edu.cn}

\begin{abstract}
We present our investigation of HI-rich `Dark' galaxiEs in Simulations (HIDES), specifically using the \textsc{hestia} and Auriga simulations in this work. 
We select galaxies that are faint ($M_g > -10$) and contain sufficient HI ($M_\mathrm{HI} > 10^5\,M_\odot$), and identify 89 such objects, only one of which is completely starless. 
Their demographics generally converge across simulations of different resolution, with $M_{200} \sim 10^{9.5}\,M_\odot$, $M_\mathrm{gas} \sim 10^{7.4}\,M_\odot$, $M_\mathrm{HI} \sim 10^{6.5}\,M_\odot$, $M_\mathrm{*} \sim 10^{5.6}\,M_\odot$, low gas metallicity, little or no current star formation, and a mean stellar age of $\sim$ 11 Gyr, and with some of them can survive in dense environments as close as $\sim$ 300 kpc from a Milky-Way mass neighbor. 
We find a large scatter in their HI density profiles and $M_\mathrm{HI} - M_\mathrm{*}$ relation, which cannot be fully explained by current halo mass or concentration, but can be attributed to ram pressure stripping in dense environments, past mergers, and stellar feedback. 
In particular, close encounters with massive halos and dense environments can reshape the HI content, which may explain the asymmetric HI map of an intriguing observed analogue, Cloud-9. 
An empirical fit, $n = 0.25 \left(d_\mathrm{MW}/{1\,\mathrm{Mpc}}\right)^{-1.4}\, \mathrm{Mpc}^{-3}$, based on their number density extended to 3.7 Mpc in constrained local volume simulations, is also provided to aid observational forecasts.
We conclude that both mass assembly history and environmental history play a crucial role in the formation and subsequent diversity of these galaxies. 

\end{abstract}


\section{Introduction}
\label{sec:intro}

The $\Lambda$CDM framework \citep[e.g.,][]{Davis1985, White&Frenk1991} has been profoundly successful in explaining a wide range of independent observations in cosmology. 
It connects the formation and evolution of dark matter halos to primordial fluctuations on microscopic scales – taking dark matter as 100 GeV weakly interacting massive particles for instance, it collapses and forms halos from Earth to galaxy cluster mass \citep[$10^{-6}-10^{15.5}\,M_\odot$,][]{Wang2020, Zheng2024a, Liu2024}. 
In this picture, massive halos ($\gtrsim 10^{9.5}\,M_\odot$, depending on redshift) act as gravitational potential wells to trap and cool gas, thereby triggering galaxy formation \citep{Rees1977, White&Rees1978}. 
However, not all halos are able to form stars, as the UV radiation released from already-formed massive galaxies heats and expels gas from small halos with shallow potential wells \citep[$\lesssim 10^{8}\,M_\odot$, as a function of time, e.g.,][]{Efstathiou1992, Klypin1999, Bullock2000, Okamoto2008, Sawala2016a, Zheng2024b}. 

In the mass regime between these two extremes (i.e., star formation and gas removal), there should be halos that have accreted gas, but have not yet, or have barely formed any stars \citep{Rees1986, Ikeuchi1986}. 
For example, \citet{Benitez-Llambay2017} (hereafter \citetalias{Benitez-Llambay2017}), using the Apostle simulations \citep{Sawala2016, Fattahi2016}, identified a population of halos $\sim 10^{8.5}-10^{9.7}\,M_\odot$, whose star formation is completely suppressed by reionization yet retains a central neutral hydrogen core, and named them RELHICs (REionization-Limited HI Clouds). 
In particular, \citetalias{Benitez-Llambay2017} developed an analytical model to predict the  inner gas structure based on halo mass, concentration, and a tight relation between gas temperature and density they found in the Apostle RELHICs. 
\citet{Rey2022}, who studied halos of a similar mass ($10^{9.1} - 10^{9.6}\,M_\odot$), reported their findings with HI-rich faint dwarfs (stellar mass $M_* \lesssim 10^6\, M_\odot$) in the EDGE simulations \citep{Rey2019, Rey2020, Agertz2020, Pontzen2021}, showing a bimodality in their cold gas mass and attributing it to the coupling between the response of cold gas to UV radiation, scatter on stellar-to-halo mass relation, and feedback-driven time evolution. The existence of such `dark' galaxies in $\Lambda$CDM simulations (i.e., either completely starless as RELHICs, or faint enough to be missed in optical bands), is further corroborated in other studies \citep[e.g.,][]{Jimenez2020, Lee2024, Doppel2025}, but there is still no consensus on their properties and the extent of diversity, particularly regarding the stellar and HI components, as they vary across different simulations. 

As a crucial examination of the $\Lambda$CDM model at small scales, enormous efforts through 21 cm observations \citep[e.g.,][]{Minchin2005, Adams2013, Cannon2015, Leisman2021, Xu2023, Kwon2025, O'Beirne2025} have been devoted to searching for these intermediate mass halos, particularly those inhabiting HI-rich `dark' galaxies and can be identified as `dark' HI clouds. 
Using the FAST radio telescope, \citet{Zhou2023} reported a candidate of such objects: Cloud-9, later followed by \citet{Karunakaran2024, Benitez-Llambay2024} with the Green Bank Telescope (GBT) and Very Large array (VLA). It is one of the most promising candidates to date – it is speculated to have an HI mass $\sim 10^6-2\times 10^6\, M_\odot$, stellar mass $\lesssim 10^5\,M_\odot$, a narrow linewidth, and an almost round shape \citep{Zhou2023, Benitez-Llambay2023} – features that align with the \citetalias{Benitez-Llambay2017} predictions. 
Several discrepancies have also been noted: for example, the close projected distance (109 kpc, \citealt{Zhou2023}) between Cloud-9 and a nearby Milky Way-sized galaxy M94, which potentially contradicts the isolated environment that \citetalias{Benitez-Llambay2017} suggested for RELHICs; a tail-like feature in the HI map, and a column density profile flatter than the original \citetalias{Benitez-Llambay2017} model in follow-up VLA observations with higher angular resolution \citep{Benitez-Llambay2024}, which indicates that Cloud-9 may be subject to perturbations or ram pressure stripping through interactions with the gas around M94. 
\citet{Benitez-Llambay2024} also proposed the possibility that Cloud-9 could be a luminous galaxy, with an upper stellar mass limit resembling that of Leo T, a nearby HI-rich faint dwarf galaxy. 
Although this limit has recently been narrowed down to $\sim 10^{3.5}-10^4\, M_\odot$ \citep{Anand2025}, the possibility cannot yet be fully excluded. 

Therefore, there is still a lot of work, both theoretical and observational, pressing to understand the nature of Cloud-9 or similar objects in the category of HI-rich `dark' galaxies, thereby probing the edge of galaxy formation as a testing ground for dark matter models and baryonic physics. 
In this serial work named `\textit{HIDES}' (HI-rich `Dark' galaxiEs in Simulations), we aim to provide a systematic study of these HI-rich `dark' galaxies (HIDEs hereafter), which remain hidden in the optical band but can be revealed through 21 cm observations, by combining cosmological/idealized simulations, analytical models, and observational implications. 

The present paper focuses on the population and diversity of such galaxies in the \textsc{hestia} and Auriga simulations \citep{Libeskind2020, Grand2017, Grand2024}, since they use the same code and baryonic physics, providing an additional test with subgrid models distinct from those of the Apostle and EDGE simulations \citep[i.e.,][]{Benitez-Llambay2017, Rey2022}. 
Particularly, we use the constrained local group environment of \textsc{hestia} to provide realistic predictions on their demographics, including abundance and radial distribution out to $3.7$ Mpc, and exploit the high resolution of Auriga to test their inner structure, formation mechanism, and numerical convergence.

The paper is organized as follows. Section \ref{sec:simulation} describes the simulation details and the definition of HIDEs. We present our main results in Section~\ref{sec:results}, and further discuss the origin of these galaxies and the scatter in their properties in Section~\ref{sec:discussion}. In Section~\ref{sec:conclusion}, we summarize and conclude our results.

\section{Simulation and Methods}
\label{sec:simulation} 
\subsection{Simulation}

The \textsc{hestia} simulation suite \citep{Libeskind2020} consists of a series of cosmological simulations focusing on the environment of the Local Group, which is constrained by the peculiar velocity observations of nearby galaxies from the CosmicFlow-2 catalogue \citep{Tully2013}. 
It adopts cosmological parameters from \citet{Planck2014p16} with $h=0.6777$, $\sigma_8=0.83$, $\Omega_\mathrm{m}=0.318$ and $\Omega_\mathrm{b}=0.048$, and is performed with the moving-mesh magnetohydrodynamics code \textsc{arepo} \citep{Springel2010, Pakmor2016}.   The subgrid physics is inherited from the Auriga galaxy formation model \citep{Grand2017}, which incorporates the key physical processes of galaxy formation and evolution, and has been proven to reproduce many properties of the Milky Way and its satellites with great numerical convergence across different scales and resolutions \citep[e.g.,][]{Grand2017, Grand2018, Marinacci2017, Simpson2018, Liao2019, Monachesi2019, Fattahi2020}. 
The Auriga model describes the interstellar medium with two phases, i.e., the cold star-forming gas and the hot surrounding gas exchanging mass with each other through radiative cooling, star formation, and feedbacks \citep{Springel2003}. For the focus of this paper, it is worth noting that this model includes gas cooling via primordial and metal species, along with self-shielding corrections against a uniform UV background, as well as star formation with a fixed density threshold of $n=0.13~\mathrm{cm}^{-3}$ for gas cells \citep{Vogelsberger2013}. The interested readers are referred to \citet{Grand2017} for more details. 

For the \textsc{hestia} highest resolution runs used in the paper \citep[i.e., the three runs, `09\_18', `17\_11', and `37\_11' as in][]{Libeskind2020}, the zoom-in regions are centred by two galaxies with properties resembling the Milky Way and M31's (e.g., virial mass, stellar mass, relative distance and velocity, etc.), covering two overlapping $3.7~\mathrm{Mpc}$ spheres at $z=0$. 
Such a large and realistically constrained volume makes \textsc{hestia} the ideal simulations for searching dark or ultra-faint galaxies nearby. 
The mass of high resolution dark matter particles is $m_\mathrm{dm}\sim 2.01\times10^5~M_\odot$, the initial mass of gas cells is $m_\mathrm{gas,\,init}\sim 3.58\times10^4~M_\odot$, and the softening length is $\epsilon=360~\mathrm{cpc}$, and is fixed at $180~\mathrm{ppc}$ after $z=1$. 

To dissect the galaxy inner structure and test the numerical convergence of our results, we use the Auriga simulations \citep{Grand2017, Grand2024} with the same recipe, smaller and non-constrained zoom-in volume but higher resolution, to resolve galaxies of our interest with more particles.\footnote{Note that the adopted cosmological parameters are slightly different (e.g., $\Omega_\mathrm{m}=0.307$), but the impact should be negligible for our purpose.} Specifically, we use one Auriga L2 run with the highest resolution and snapshot output frequency to present the gas profile and evolution history, and all six Auriga L3 runs with more samples for the convergence test. The corresponding parameters are listed as follows: $m_\mathrm{dm,~L2~(L3)}=2.09\times10^3~(1.67\times10^4)\,M_\odot$, $m_\mathrm{gas,~init,~L2~(L3)}=3.87\times10^2~(3.10\times10^3)\,M_\odot$, $\epsilon_\mathrm{DM,~L2~(L3)}=184~(369)~\mathrm{cpc}$ before $z=1$ and is fixed at $92~(184)~\mathrm{ppc}$ after, $\epsilon_\mathrm{gas,~L2~(L3)}$ is scaled by the mean radius of the cell, with a limit between $184~(369)~\mathrm{cpc}$ and $92~(922)~\mathrm{ppc}$. 

The \textsc{hestia} and Auriga simulations adopt the friends-of-friends \citep[FOF,][]{Davis1985} and \textsc{subfind} algorithm \citep{Springel2001, Dolag2009} to identify halos and subhalos, and the \textsc{LHaloTree} \citep{Springel2005} to build merger trees; \textsc{hestia} additionally provides the halo catalog identified with \textsc{ahf} \citep{Knollmann2009}. For easier comparison, we follow the halo definition of \citetalias{Benitez-Llambay2017}, using central halos and their properties within the virial radius, $r_\mathrm{200}$ (i.e., the radius within which the mean matter density is equal to 200 times the cosmic critical density) unless otherwise specified. 

To avoid numerical contamination from low resolution particles, we apply a spatial cut in the \textsc{hestia} runs, that is, only consider the galaxies inside two 3.7 $\mathrm{Mpc}$ spheres around the Milky Way and M31. To increase the sample size in the Auriga runs, we use a less strict selection criterion, considering the galaxies without low resolution particles within 5 times their own virial radius.

\subsection{Definition of HIDEs (HI-rich `dark' galaxies)}

\begin{figure*}
    \centering
    \includegraphics[width=1.0\columnwidth]{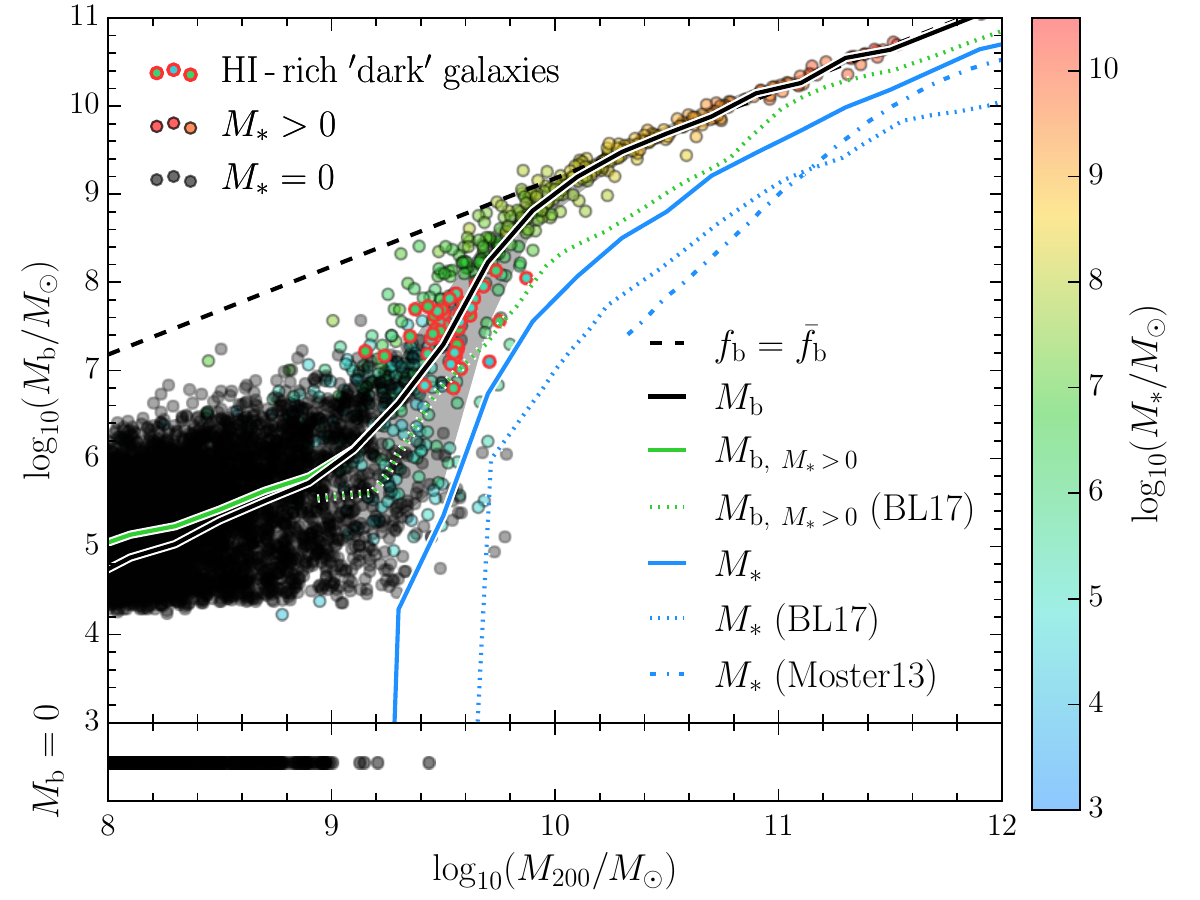}
    \includegraphics[width=1.0\columnwidth]{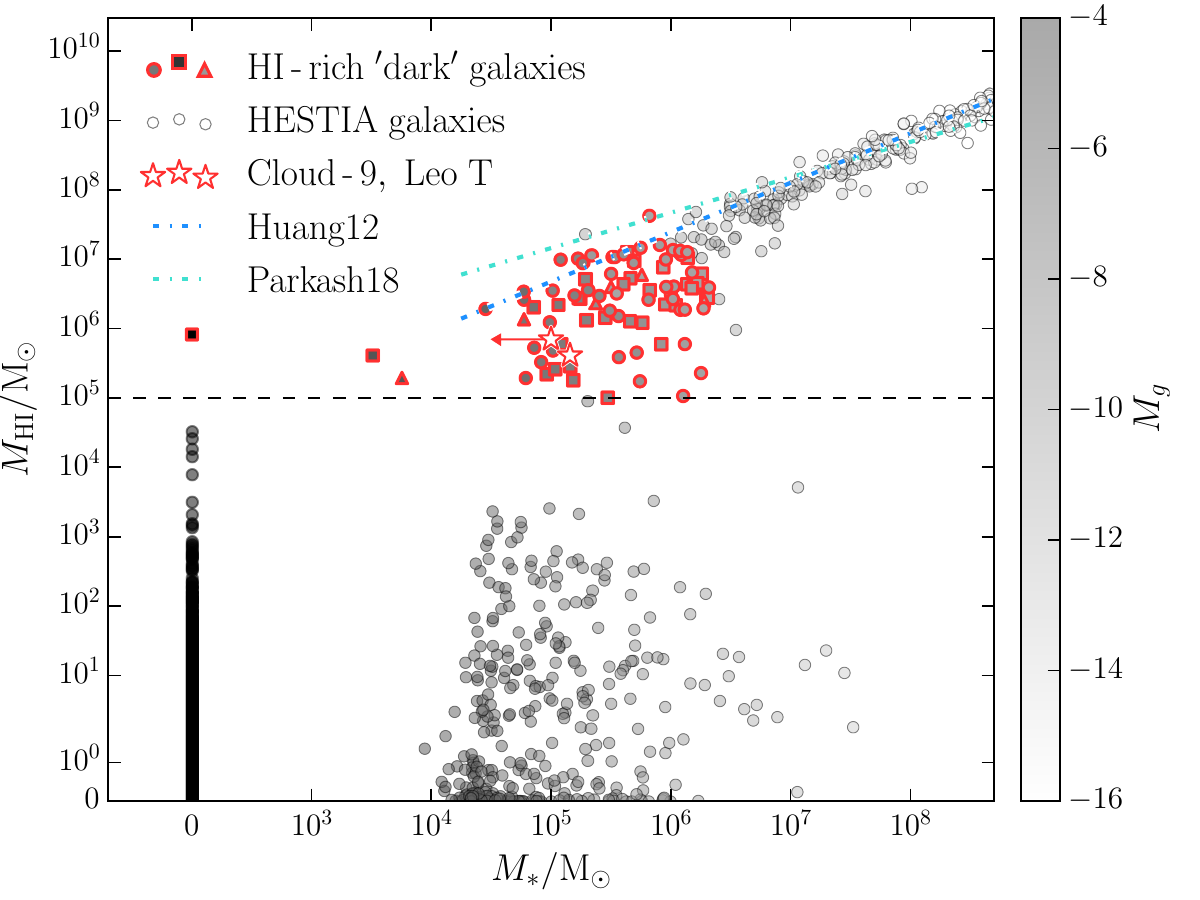}
    
	\vspace{-0.1cm}
	\caption{The baryon mass vs. halo mass relation (left panel) and HI mass vs. stellar mass relation (right panel) of galaxies in the \textsc{hestia} simulation. 
    Left: colored dots: halos colored by their stellar mass, black dots: completely starless halos; black solid line: median baryon mass; black dashed line: cosmic mean baryonic fraction; green lines: the median baryonic mass of halos that contain stars (solid: \textsc{hestia}, dotted: \citetalias{Benitez-Llambay2017}); blue lines: median stellar mass (solid: \textsc{hestia}, dotted: \citetalias{Benitez-Llambay2017}, dash-dotted: observational results of \citealt{Moster2013} via abundance matching). 
    Right: gray dots: halos colored with the $g$-band magnitude; black dots: completely starless halos (i.e., $M_*=0$); red stars: observational results of Cloud-9 and Leo T, dash-dotted lines: extrapolated $M_\mathrm{HI}$ - $M_*$ fits obtained from observations of more massive galaxies. 
    In \textsc{hestia}, We do not find completely starless halos with HI masses comparable to Cloud-9 and Leo T, therefore we select HIDEs (HI-rich `dark' halos, red-circled dots) as our sample instead, which have a great amount of HI ($M_\mathrm{HI}>10^5\,M_\odot$) and are faint enough to possibly `hide' themselves from wide-field optical surveys ($M_g > -10$); HIDEs selected with the same criteria from Auriga L3 and Auriga L2 are overplotted (red-circled squares and triangles, respectively) for comparison.} 
	\label{fig:fig1}
\end{figure*}

To study the nature of Cloud-9 and Leo T with simulations, we need to identify their counterparts in simulated galaxies, which requires a criterion to define such a population. 
In the left panel of Fig. \ref{fig:fig1}, we first plot all central galaxies on the baryon mass – halo mass plane, comparing to the Figure 1 of \citetalias{Benitez-Llambay2017}. The colored dots represents those halos with stellar particles within $r_{200}$, while the black dots representing those being completely starless. 
The median values are plotted with solid lines – the green solid line representing halos with stellar particles is mostly overlapped with the black solid line representing all samples, 
and is above the green dotted line adapted from \citetalias{Benitez-Llambay2017} by $\sim$ 0.5 to 1 dex between $10^{9}$ and $10^{10.5}\,M_\odot$. 
The halos with mass above $10^{10}\,M_\odot$ generally have a baryonic fraction close to the cosmic value $\bar{f_\mathrm{b}}$, and those smaller halos smoothly transit to around $0.01\bar{f_\mathrm{b}}$ at halo mass $\gtrsim 10^9~\mathrm{M}_\odot$. 
Meanwhile, the blue solid lines representing stellar mass is also always above \citetalias{Benitez-Llambay2017} and extends to lower mass end before the median stellar mass drops to zero. 
This indicates the gas accretion and star formation are more efficient in \textsc{hestia} than in Apostle, which aligns with the findings of \citet{Kelly2022}, who reported higher stellar masses in Auriga galaxies compared to their Apostle counterparts, although at a much higher halo mass range than the focus of our study. 
This difference is likely due to different subgrid recipes, such as the setup of star formation threshold (parametrized as a metallicity dependent value in the Apostle recipe, which would exponentially grow in low-metallicity systems like dark halos and prevent star formation inside) and stellar feedback. 
Another important difference we find, is the baryon masses in completely starless halos spread as a continuum, aligning with the Fig. 11 of \citet{Pereira-Wilson2023}, but different from those in \citetalias{Benitez-Llambay2017} which can be easily divided into two distinct populations (i.e., gas-poor COSWEBS, and gas-rich RELHICs with a tight relation between baryon and halo mass).

In the right panel of Fig. \ref{fig:fig1}, we further plot all central halos on the HI mass\footnote{Note that the HI mass is calculated by summing up the HI mass in each gas particles (based on the hydrogen fraction and gas temperature), thus the resolution should be considered with the gas particle number ($M_\mathrm{gas}/m_\mathrm{gas,\, init}$) rather than $M_\mathrm{HI}/m_\mathrm{gas,\, init}$. } – stellar mass plane, comparing to Cloud-9 and Leo T shown with red stars. 
To select HI-rich counterparts, we first adopt a HI-mass lower limit of $10^5\,M_\odot$, which is around one order of magnitude below Cloud-9 or Leo T's, so as to ensure the resemblance. 
Notably, we do not find any starless halo with HI mass above this threshold in \textsc{hestia} and only one such object in Auriga L3, indicating they are very rare populations under the Auriga or \textsc{hestia} recipe, and the transition out of starless cloud is rapid as soon as it accumulates enough HI mass, which corroborates our previous findings that the \textsc{hestia} low-mass halos are relatively easier to form stars. 
In this case, to find simulation counterparts, we turn to galaxies that are faint enough to be missed in optical surveys, and set a lower limit for the $g$-band magnitude of $-10$, which corresponds to a stellar mass limit of $\lesssim 10^6\,M_\odot$. 

Finally, we select 49 samples in 3 \textsc{hestia} runs, 34 samples in six Auriga L3 runs, and 6 samples in one Auriga L2 run (i.e., the red-circled dots, squares, and triangles in the right panel of Fig. \ref{fig:fig1}, respectively), and define them as the HI-rich `dark' galaxies in this study. In this way, we include those samples with optical counterparts that might be too faint to detect in wide galaxy surveys. 
We find that although they generally align with the extrapolation of observational fits obtained at high mass end \citep[e.g.,][]{Huang2012, Parkash2018}, but with much larger scatters comparing to more massive galaxies with $M_* > 10^{6.5}\,M_\odot$. This diversity on the $M_\mathrm{HI}-M_*$ plane, along with the bimodal distribution of HI mass,  is further confirmed in Auriga L2 and L3 runs with higher resolution and aligns with the distribution of cold gas mass that \citet{Rey2022} found for faint dwarf galaxies in the EDGE simulation. 

\begin{figure*}
    \centering
    \includegraphics[width=2.0\columnwidth]{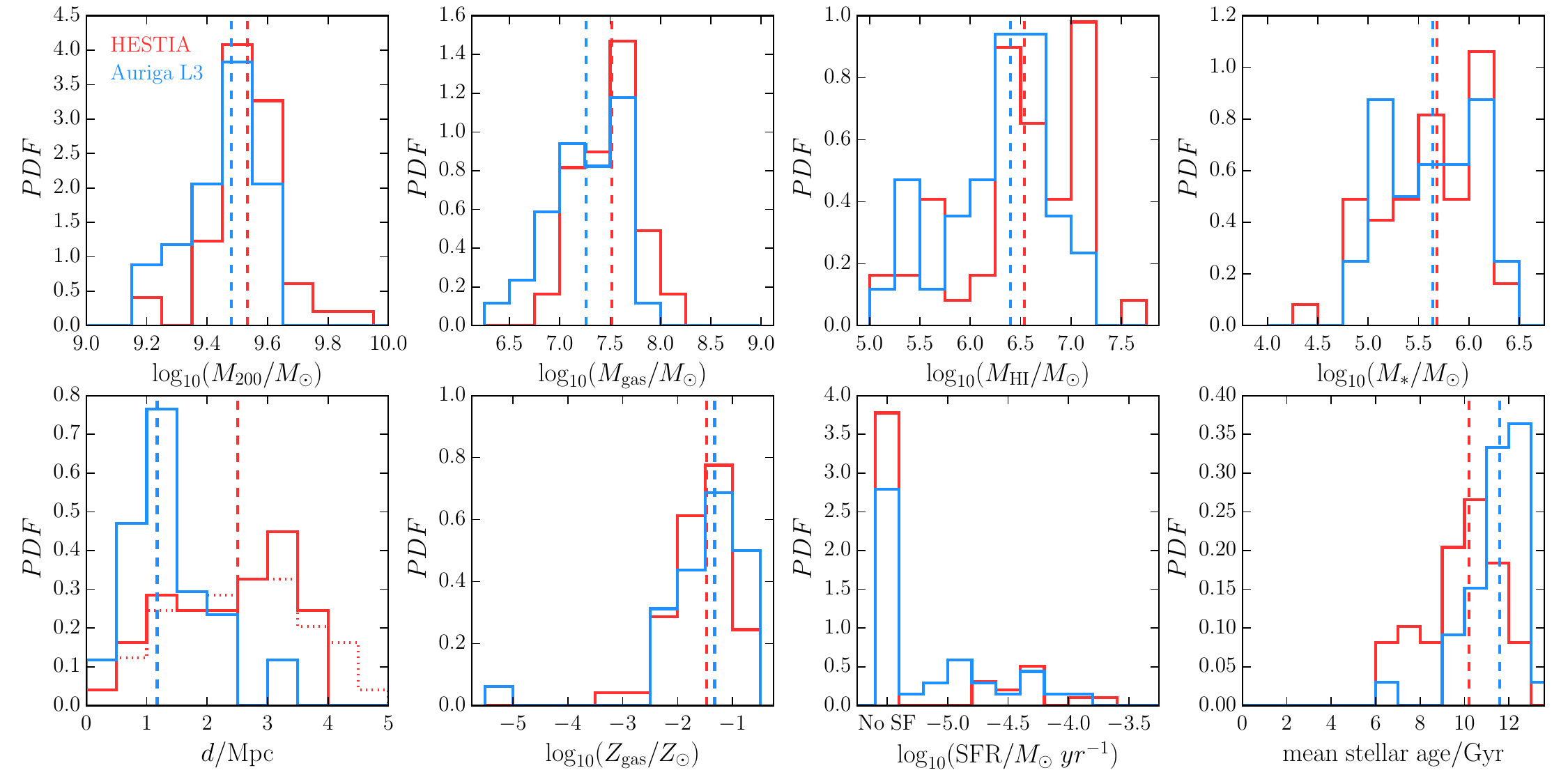}
    
	\vspace{-0.1cm}
	\caption{Histograms of the general properties of HIDEs in \textsc{hestia} (red) and Auriga L3 (blue), with the vertical dashed lines showing the median values. In the bottom left panel, the red dotted line shows the distribution of distances to the Milky Way, which differs from the minimum distances to either M31 or the Milky Way shown with red solid lines, and there is no need to distinguish these definitions in Auriga as there is only one host galaxy. 
    HESITA and Auriga HIDEs have largely converged properties, with $M_{200} \sim 10^{9.5}\,M_\odot$, $M_\mathrm{gas} \sim 10^{7.4}\,M_\odot$, $M_\mathrm{HI} \sim 10^{6.5}\,M_\odot$, $M_\mathrm{*} \sim 10^{5.6}\,M_\odot$, low metallicity gas due to the low stellar mass, low or no star formation at present, and old mean stellar age $\sim 11$ Gyr. The difference in distance distribution is mainly due to the different volume sizes of two simulation suites. }

    \label{fig:fig2}
\end{figure*}

\section{The Population and Diversity of HIDEs (HI-rich `Dark' Galaxies)} \label{sec:results}

\subsection{General Properties}

In Fig. \ref{fig:fig2}, we show the histograms of general properties of our HIDEs sample in \textsc{hestia} (red lines), and further use the Auriga L3 samples (blue lines) to test the numerical convergence. 
In the first row, we show the distributions of the masses of different components: the median total, gas, HI, and stellar mass of the HIDEs in \textsc{hestia} are $10^{9.53}$, $10^{7.52}$, $10^{6.54}$, $10^{5.68}\,M_\odot$ respectively, while the corresponding values in Auriga L3 are $10^{9.48}$, $10^{7.26}$, $10^{6.40}$, $10^{5.64}\,M_\odot$. 
The total mass of the HIDEs we find here, have a halo mass between $10^{9.15}$ and $10^{9.87}\,M_\odot$ ($10^{9.22}$ and $10^{9.64}\,M_\odot$ in Auriga L3), which lies at the high mass end of the RELHICs identified in \citetalias{Benitez-Llambay2017}, indicating they might be similar objects in different simulations, but formed stars due to different designs in subgrid physics. 
The median HI and stellar mass are close between \textsc{hestia} and Auriga L3 runs, with the values in Auriga being slightly smaller (0.14 and 0.04 dex), while the difference in median gas mass is larger (0.26 dex) – this may partially due to the slightly smaller total mass (0.05 dex), and another possible cause might be the denser environment, since Auriga runs cover much smaller volumes closer to Milky Way. 
In the bottom left panel, we show the distance distribution of these galaxies, note that there are two `host' galaxies (i.e., M31 and Milky Way) is \textsc{hestia}, we adopt two definitions of distance: the red solid line shows the minimum distance to M31 and the Milky Way, and the red dotted line shows the distance to the Milky Way (i.e., the smaller host). And the distance of Auriga L3 HIDEs, missing the volume coverage at larger radii, is indeed much smaller than \textsc{hestia} HIDEs. 
The gas is suggested to be metal poor in both simulations ($10^{-1.47}$ and $10^{-1.33}\,Z_\odot$) and mostly not star-forming, aligning with fact that there are only a few and old star population (10.19 and 11.58 Gyr) inside. 
In brief, we only find minor difference between the \textsc{hestia} and Auriga L3 HIDEs, suggesting the general properties are mostly converged at the \textsc{hestia} resolution.

\begin{figure*}
    \centering
    \includegraphics[width=0.868\columnwidth]{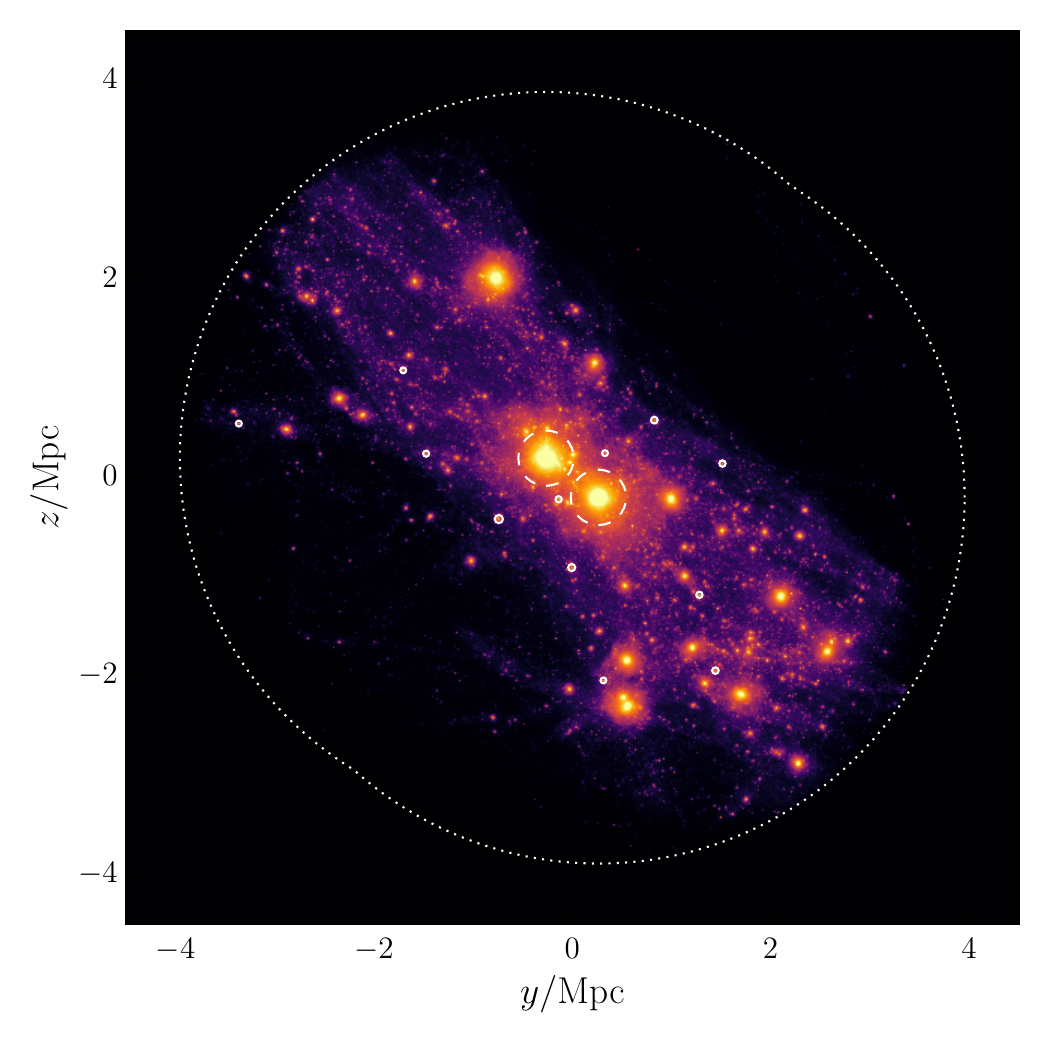}
    \includegraphics[width=1.132\columnwidth]{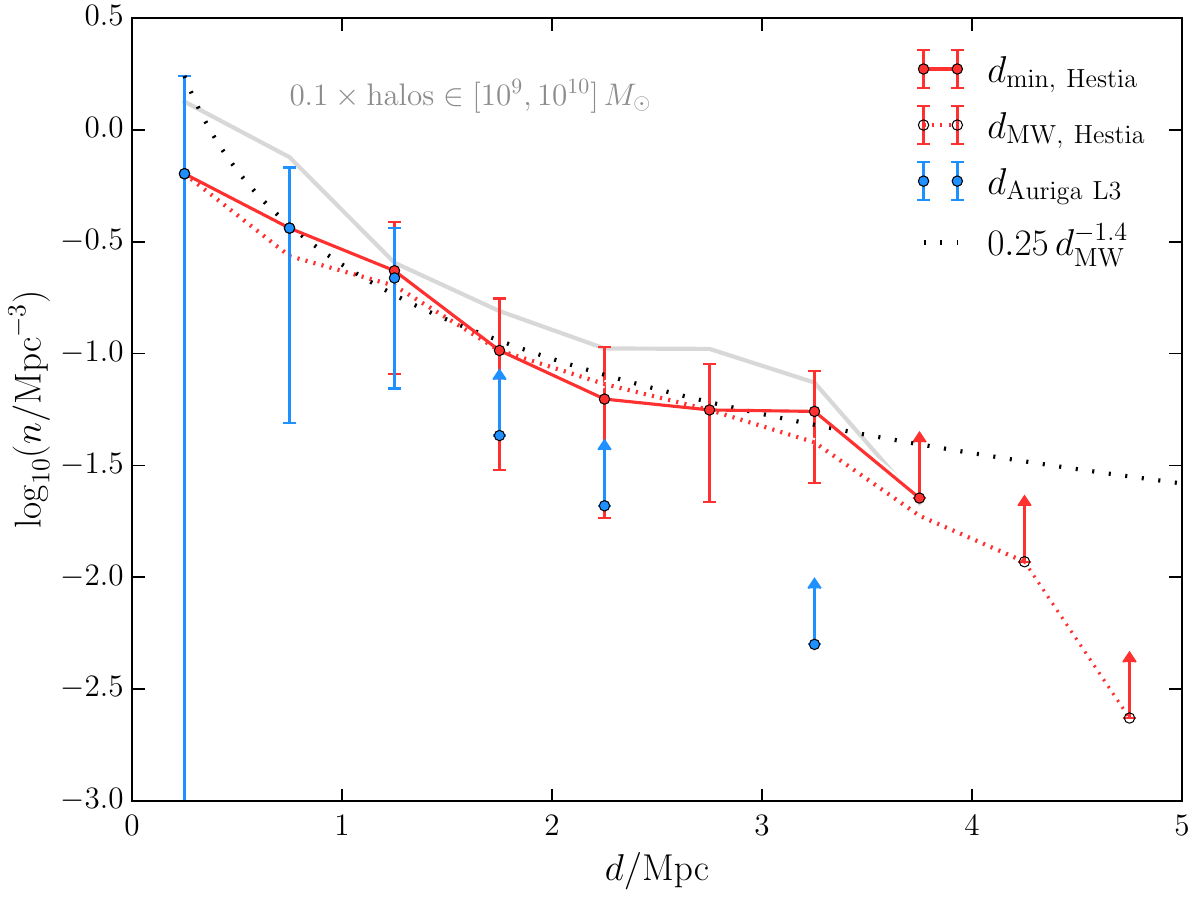}
    
	\vspace{-0.1cm}
	\caption{The spatial distribution of HIDEs. Left: projected dark matter density field, with solid, dashed, and dotted circles representing HIDEs, M31 and the Milky Way, the boundary of zoom-in region respectively.
    Right: number density of HIDEs as a function of radius, averaged among different realizations. Red solid line: the smaller distance to either M31 and the Milky Way in \textsc{hestia}; red dotted line: the distance to the Milky Way in \textsc{hestia}; blue line: the distance to the central host galaxy in Auriga L3; error bars: Poisson errors; error with arrows: lower limit due to incomplete volumes; gray line: 0.1 times the number density of halo in the relevant mass range $[10^9,\,10^{10}]\, M_\odot$. The radial distribution ca be fitted with a simple power law function (black dotted line): $n = 0.25 \left(d_\mathrm{MW}/1\,\mathrm{Mpc}\right)^{-1.4}\mathrm{Mpc}^{-3}$. }
	\label{fig:fig3}
\end{figure*}

\subsection{Abundance and Spatial Distribution}

\begin{figure*}
    \centering
    \includegraphics[width=2.0\columnwidth]{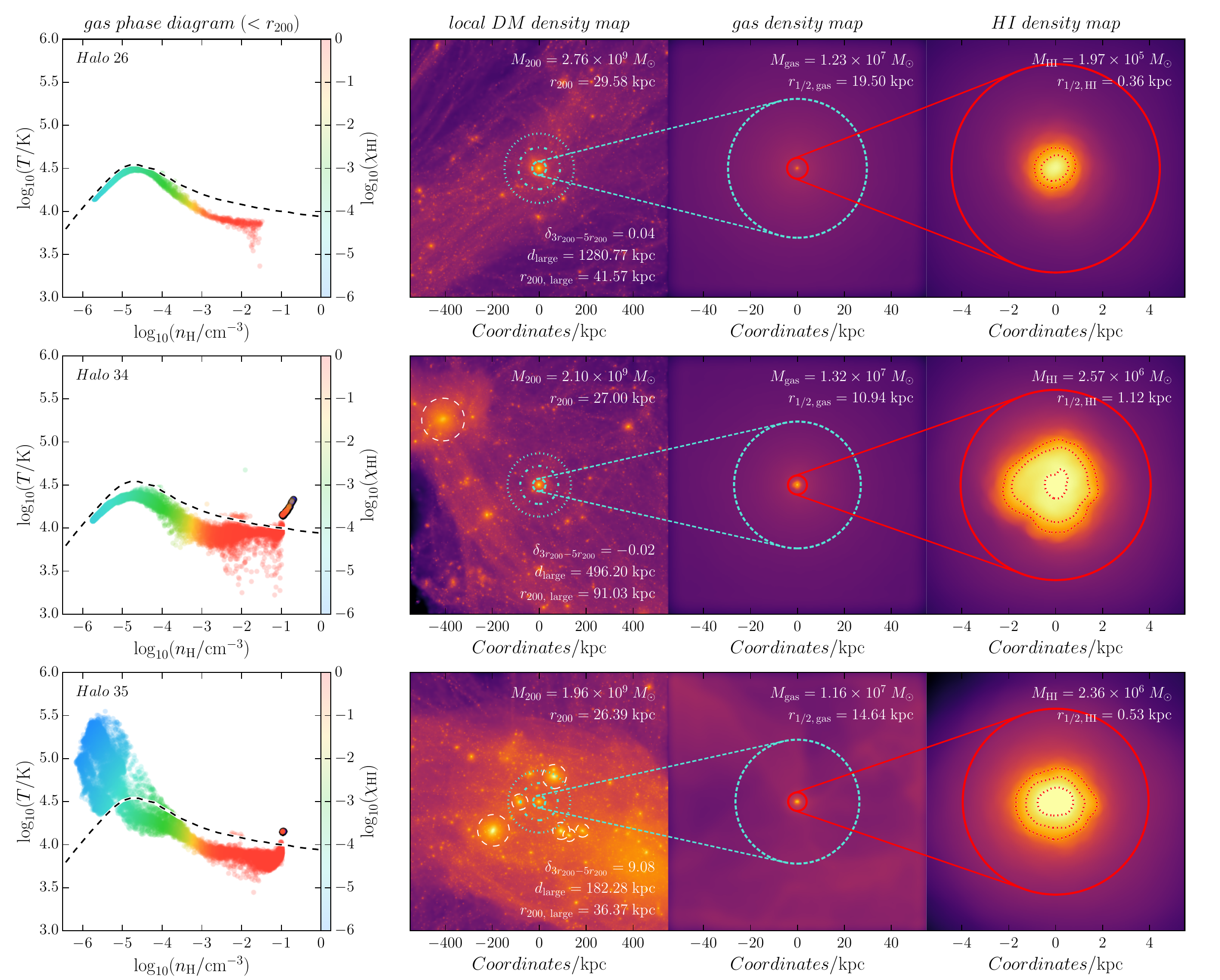}
    
	\vspace{-0.1cm}
	\caption{Gas phase diagrams and projected density field, with each row representing one case study (Halo 26, 34, 35) at $z=0$. First column: gas temperature ($T$) vs. hydrogen number density ($n_\mathrm{H}$) of all gas particles inside the halo virial radius. The color of the dots represents the HI mass fraction of gas particles; dots on the right side with black edges represent gas particles that meet the star formation criteria. Dashed lines: the $T-n_\mathrm{H}$ relation fitted in \citetalias{Benitez-Llambay2017} for Apostle RELHICs. Second column: local dark matter density map, with dashed, dash-dotted, and dotted cyan circles representing the 1, 3, and 5 times the virial radius respectively, and the environmental dark matter overdensity $\delta_\mathrm{3r_{200}-5r_{200}}$ is measured in the shell between the latter two. White dashed circles indicate nearby more massive halos, and the distance and radius of the nearest one are marked as $d_\mathrm{large}$ and $r_\mathrm{200,\,large}$. Third column: zoom-in view of gas density map. Fourth column: further zoom-in view of HI density map, with red solid circles marking 0.15$r_{200}$ as a reference, and red dotted circles marking the contour of HI column density $N_\mathrm{HI} = 10^{18},\ 10^{19},$ and $10^{20}\ \mathrm{cm}^{-2}$ from outside toward center. }
	\label{fig:fig4}
\end{figure*}

We show the projected dark matter density field of one \textsc{hestia} run in the left panel of Fig. \ref{fig:fig3}, with the HIDEs marked as solid white circles. It shows that the HIDEs mostly reside in the local sheet as other halos do, without apparent dependence on local dark matter density. A zoom-in view of dark matter density field around 3 samples from the Auriga L2 run can be found in Fig. \ref{fig:fig4}. 
The radial distribution of HIDEs are shown in the right panel of Fig. \ref{fig:fig3}. Following our previous notation, we use red solid line to show the smaller distance to M31 and to the Milky Way, and use red dotted line to show the distance to the Milky Way. The error bars show the Poisson errors in each radial bin, with those with arrows suggesting incomplete volume. 
It shows that at overlapping radii ($<$ 1.5 Mpc), the HIDEs in \textsc{hestia} and Auriga L3 follow a similar radial distribution. 
The \textsc{hestia} samples further extend the complete coverage to $3.7$ Mpc, enabling us to generate a fit\footnote{To extend the applicability of our results to different limits on $g$-band magnitude (i.e., $M_{g,\, \mathrm{limit}}$), we also generate a fit with $M_{g,\, \mathrm{limit}}$ as a variable (while fixing the limit of $M_\mathrm{HI} > 10^5\,M_\odot$): $n = 0.25 \left(d_\mathrm{MW}/{1\,\mathrm{Mpc}}\right)^{-1.4}\, 10^{-0.125(M_{g,\,\mathrm{limit}}+10)} \, \mathrm{Mpc}^{-3}$. } on the number density, with 
\begin{equation}
    n = 0.25 \left(\dfrac{d_\mathrm{MW}}{1\,\mathrm{Mpc}}\right)^{-1.4}\mathrm{Mpc}^{-3}. 
\end{equation}
The nearest HIDEs has a distance of 305 kpc (368 kpc in Auriga L3), which more or less coincides with the distance of Leo T \citep[$\sim$~409 kpc,][]{Clementini2012}, indicating that HIDEs, although occupy a slightly lower fraction in the halo population at such distance (as shown in Fig. \ref{fig:fig3}), can possibly survive in relatively dense environment\footnote{The environment is, later on, quantified with the dark matter overdensity within the shell of $3-5$ times of the virial radius, $\delta_{3r_{200}-5r_{200}}$ – for this \textsc{hestia} case, it has a $\delta_{3r_{200}-5r_{200}}=36.05$, indicating a very dense local environment. }
, which is slightly different from the RELHICs in \citetalias{Benitez-Llambay2017} found to be mostly isolated. 

\subsection{Variant Gas Inner structures under the Impacts of Star Formation and Environment}

\begin{figure*}
    \centering
    \includegraphics[width=2.0\columnwidth]{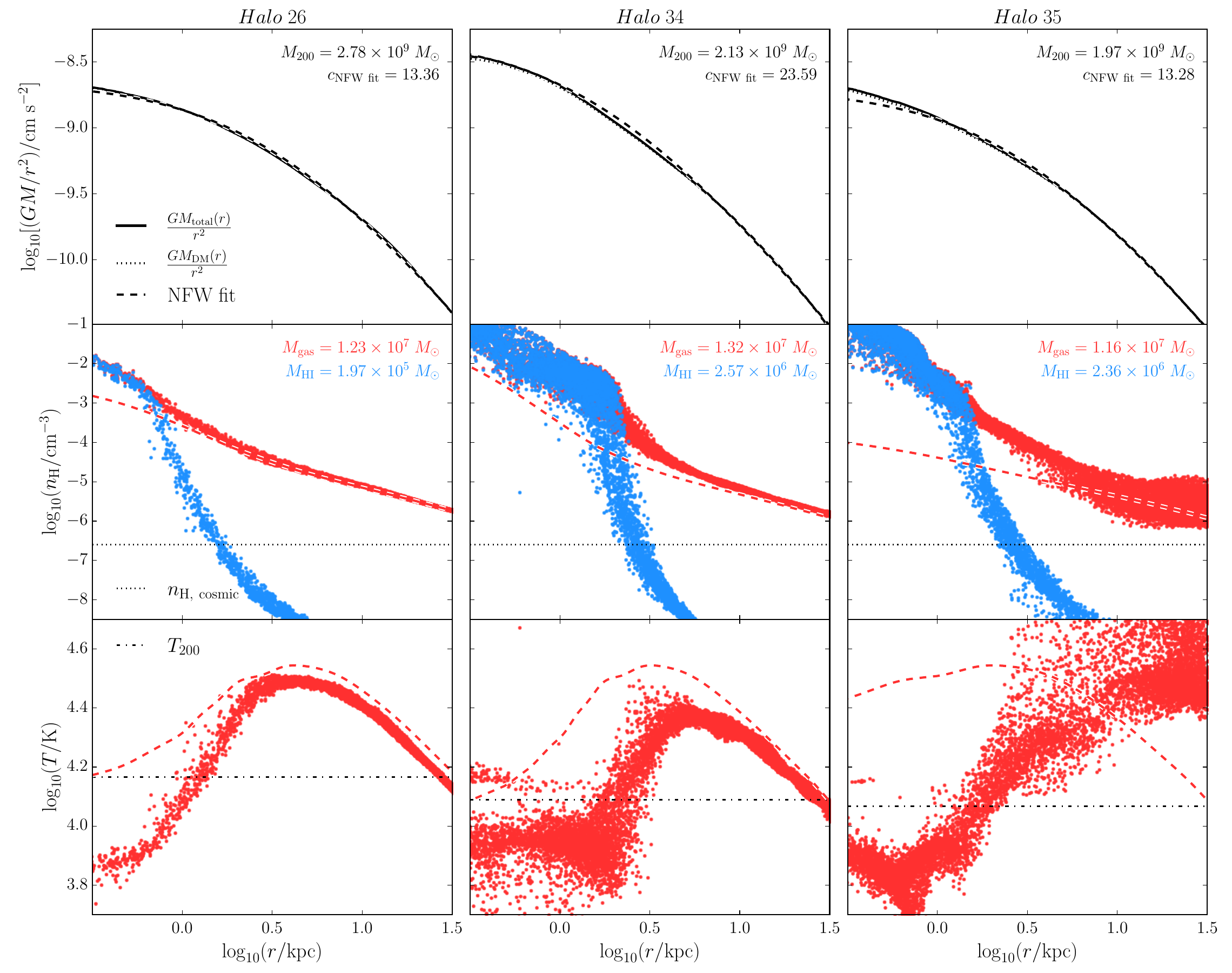}
    
	\vspace{-0.1cm}
	\caption{Inner structures of 3 HIDEs, with each column representing one case study (the same cases as Fig. \ref{fig:fig4}). 
    First row: the profiles of gravitational acceleration (solid lines: acceleration from total mass; dotted lines: acceleration from dark matter mass; dashed lines: acceleration of halos with fitted NFW profile). 
    Second and third rows: red (blue) dots: gas (HI) density and temperature of each gas particle; red dashed lines: gas profile predicted by the \citetalias{Benitez-Llambay2017} model based on the halo mass and fitted concentration; black dotted lines: cosmic mean hydrogen density; black dash-dotted lines: halo virial temperature. }
	\label{fig:fig5}
\end{figure*}

To understand the impact of environment, stars and possible star formation activity on these galaxies while minimizing numerical concerns, we pick 3 HIDEs (Halo 26, 34, 35, respectively on each row) with similar mass from the highest resolution run, i.e., Auriga L2\footnote{Similar cases can be found in \textsc{hestia} but are not shown for simplicity. }, and plot their local matter density field\footnote{A public code \textsc{Py-SPHViewer} \citep{alejandro_benitez_llambay_2015_21703} is used here to smooth the density map. } and gas phase diagram in Fig. \ref{fig:fig4} and gas inner profiles in Fig. \ref{fig:fig5}. 

In the first row, we show a galaxy (Halo 26) with a tight relation between hydrogen atomic density $n_\mathrm{H}$ and gas temperature $T$ on the gas phase diagram, which is a crucial feature \citetalias{Benitez-Llambay2017} found with all the RELHICs in Apostle and used in their analytical model on gas profile. 
Compared to the relation in \citetalias{Benitez-Llambay2017} (i.e., the black dashed line), Halo 26 has a similar power-law shaped $n_\mathrm{H}-T$ relation in the outskirts ($n_\mathrm{H} \lesssim 10^{-5}\,\mathrm{cm}^{-3}$, reader can refer to \citealt{McQuinn2016} for a review for this relation), but lower temperatures in the centre ($n_\mathrm{H} \gtrsim 10^{-3}\,\mathrm{cm}^{-3}$) by $\sim$ 0.25 dex. 
As noted by \citetalias{Benitez-Llambay2017}, as gas gets denser ($n_\mathrm{H} \gtrsim 10^{-4.8}\,\mathrm{cm}^{-3}$), radiative cooling becomes more important and reaches equilibrium with the ionizing background, leading to a lower temperature in the halo centre. 
Thus, this difference likely arises from the different numerical implementations on cooling and photoheating, for example, the different cooling tables used in the Apostle and Auriga – Apostle, using the EAGLE recipe \citep{Schaye2015}, adopts the \citet{Haardt2001} model for the uniform UV background, while Auriga instead uses the \citet{Faucher-Giguere2009} UV background model additionally including a self-shielding correction, which becomes crucial in the same density range \citep[$\gtrsim 10^{-3}-10^{-2}\,\mathrm{cm}^{-3}$, e.g.,][]{Rahmati2013, Vogelsberger2013} exactly where we find the temperature difference. 
In the right panels, we show the density maps of dark matter, gas, and HI around this galaxy. We find Halo 26 resides in a filament-shaped structure but has a local density (measured between $3r_{200}-5r_{200}$) almost equal to the cosmic mean value. 
Its HI condenses at the halo centre in an almost spherical shape, with a half HI mass radius $r_\mathrm{1/2,\,HI}=0.36\,\mathrm{kpc}$. 
In the left column of Fig. \ref{fig:fig5}, we show the inner profiles of this halo and test the applicability of the \citetalias{Benitez-Llambay2017} analytical model to the HIDEs in our study. 
In the first row, the gravity acceleration profile is well aligned with the corresponding fitted NFW profile \citep{Navarro1996}, indicating these halos preserve a almost pristine dark matter profile which barely affected by baryon physics: the difference between the total matter and dark matter density is marginal, as baryons only contribute to a small mass fraction ($\sim$ 1\%). 
Using the halo mass and concentration obtained from the fit, we apply the \citetalias{Benitez-Llambay2017} analytical model to derive the gas density and temperature profile. 
We find this model is able to accurately predict the profiles in the outskirts, while it underestimates the central gas density and overestimates the central gas temperature, where $n_\mathrm{H} \gtrsim 10^{-3}\,\mathrm{cm}^{-3}$ and the deviation in the phase diagram takes effect. 
We have tested with a $n_\mathrm{H} - T$ relation calibrated with this halo's phase diagram, and found that the predicted profiles agree better with the simulation. 

The case of Halo 34 is shown in the second row of Fig. \ref{fig:fig4} and the second column of Fig. \ref{fig:fig5}. Having formed much more stars ($M_* = 1.86 \times 10^6\,M_\odot$) compared to Halo 26 ($M_* = 5.64 \times 10^3\,M_\odot$) and still star-forming, it has higher gas metallicity ($0.18\,Z_\odot$ compared to $1.1\times10^{-3}\,Z_\odot$ in Halo 26), and the corresponding metal cooling may explains the lower peak temperature at $\sim$ 4 kpc. 
The central gas temperature is relatively higher and more dispersed, almost isothermal within $\sim$ 2.5 kpc, possibly due to stellar feedback as the Halo 34 has a relatively higher star formation rate (see Fig. \ref{fig:fig6}). 
Interestingly, such a flat central temperature profile causes a shallower central density slope which coincides with Cloud-9's observation \citep{Benitez-Llambay2024}, but note that it lies close to the temperature floor set by atomic hydrogen cooling – we will address the caveats in Section \ref{sec:discussion}. 

Halo 35 is another intriguing case that has HI mass more than an order of magnitude higher than Halo 26, despite its smaller halo mass and almost identical concentration. 
It instead, shows a higher median and a larger scatter in the temperature of outskirt gas. 
Possible causes include ram pressure and stellar feedback from itself or from nearby massive galaxies. 
Based on Halo 35's properties, the ram pressure it experiences in a dense environment ($\delta_{3r_{200}-5r_{200}}=9.08$) is the more likely reason, because: i) Halo 34, which has a star-formation rate ten times higher but resides in an isolated environment, still shows a power-law temperature–density relation in the outskirts; ii) it is at least five times of the virial radius away from nearby massive halos (e.g., $d_\mathrm{large}=182.28\,\mathrm{kpc}$, $r_\mathrm{200,\,large}=36.37\,\mathrm{kpc}$). 
Such external pressure and temperature-density relation that deviate from the simplified assumption of the \citetalias{Benitez-Llambay2017} model, may also explain why Halo 35 has a central gas density far exceeding its prediction. 

As a summary of the inner structures of these three cases, we find the gas density and temperature profiles strongly depend on the distribution of gas particles on the phase diagram, which can be affected by the local environment, star formation activities, and the subgrid recipe related to cooling and photoheating.

\section{The Origin of HIDEs and Their Diversity}
\label{sec:discussion}

\begin{figure*}
    \centering
    \includegraphics[width=2.0\columnwidth]{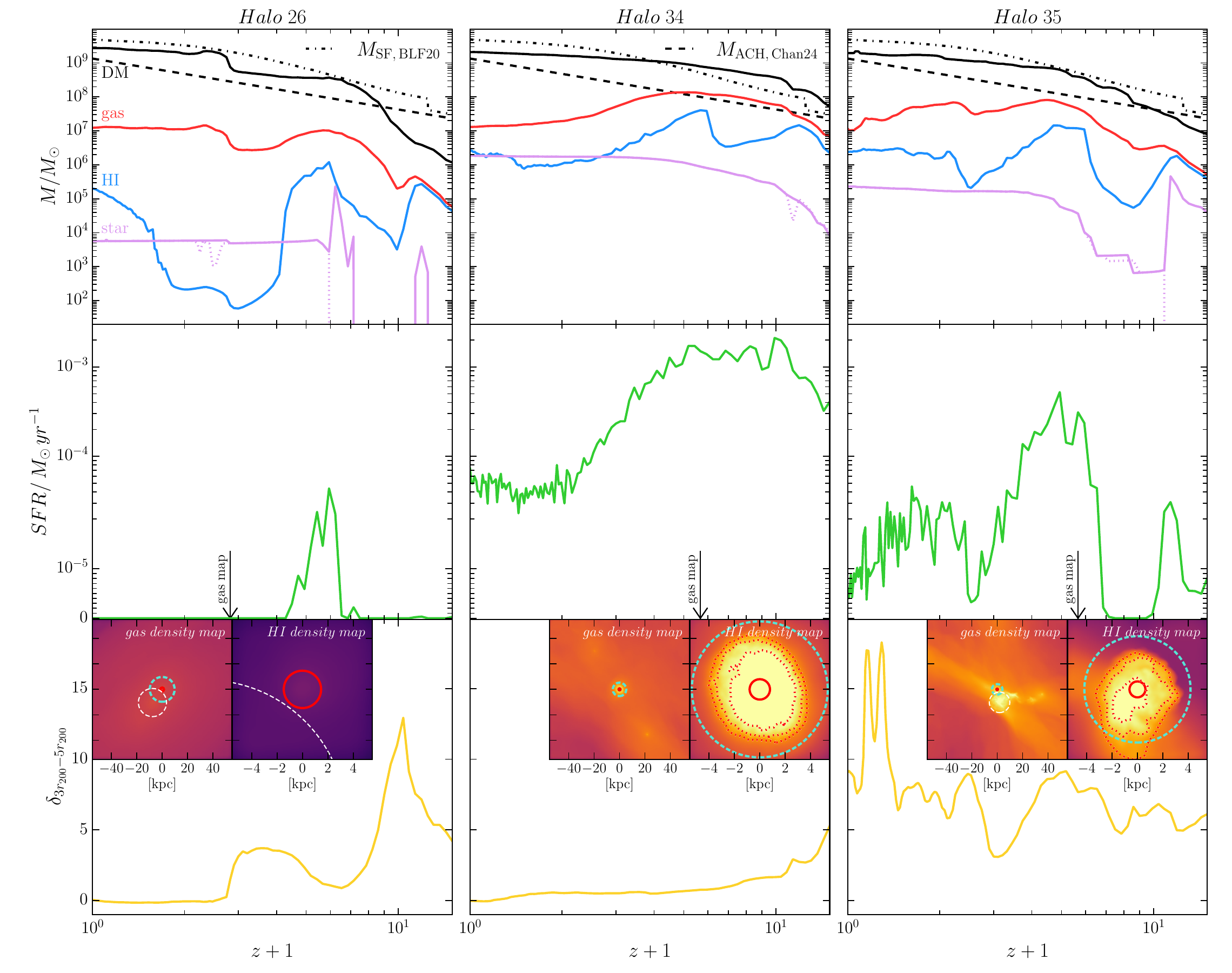}
    
	\vspace{-0.1cm}
	\caption{Evolution histories of 3 HIDEs, with each column representing one case study (the same cases as Fig. \ref{fig:fig4}). First row: solid lines: mass assembly history (within $r_{200}$) of different components (black: dark matter, red: gas, blue: HI, purple: star); purple dotted lines: stellar mass in the central subhalo; black dash-dotted and dashed lines: theoretical prediction on the halo mass corresponding to star formation ($M_\mathrm{SF,\ BLF20}$) and gas atomic cooling ($M_\mathrm{ACH,\ Chan24}$). 
    Second and third rows: history of star formation rate $SFR$ and environmental dark matter overdensity $\delta_{3r_{200}-5r_{200}}$, with the arrows representing the redshift at which the inset panel in the third row are plotted. 
    Inset panels: gas and HI density maps of the corresponding halos (Halo 26: the snapshot before a major merger; Halo 34: the moment it reached its peak HI mass; Halo 35: an aspherical shape of HI map, when it encountered a massive halo in a filament-like structure). }
	\label{fig:fig6}
\end{figure*}

As HIDEs shape the framework of galaxy formation by constraining the halo mass regime that can accrete cold gas and form stars, their origin is naturally tied to their evolutionary histories. Therefore, we study the histories of the same three HIDEs in Fig. \ref{fig:fig6}. 
On the first row, we trace the mass accretion history of the same three HIDEs, and compare with the critical mass for atomic cooling and star formation: \citet{Chan2024} defined $M_\mathrm{ACH,\,Chan24}$ as the halo mass to reach the virial temperature triggering hydrogen atomic cooling; \citet{Benitez-Llambay&Frenk2020} defined $M_\mathrm{SF,\,BLF20}$ as the mass to trigger star formation, reader can refer to those papers for detailed calculation. 
The second and third rows show the history of star formation rate and environmental dark matter overdensity respectively. 

We find all three cases here have a halo mass between $M_\mathrm{ACH,\,Chan24}$ and $M_\mathrm{SF,\,BLF20}$ for most of the time, reaffirming that HIDEs are massive enough to trigger radiative cooling, but only marginally so for significant star formation, as shown in the stellar mass (first row, purple lines) and star formation rate history (second row, green lines). 
Note that here we count all particles within the halo’s virial radius, following our previous definition, which may include particles belonging to subhalos rather than the central halo we focus on. 
Therefore, we also traced the central subhalo masses and found that the difference was generally negligible, especially for components other than stars\footnote{The results for these other components are therefore not shown in the figure.}.
The purple dotted lines in Fig. \ref{fig:fig6} show the evolution of stellar mass inside central halos — the difference at low redshifts is generally insignificant, while at earlier times the stellar mass in central halos dropped to zero for Halo 26 (at $z \sim 5$) and Halo 35 (at $z \sim 10$), which explains their unusual fluctuations in stellar mass within the virial radius (solid purple lines) caused by fly-by satellites. 
Thus, these two halos were completely starless halos before this encounter, implying the instability it induced might be the initial perturbation that broke the hydrostatic equilibrium and triggered star formation in the central halos.  

Generally, we find multiple stages in the HI mass history: 
i) HI mass closely traces the gas accretion at $z \gtrsim 10$ as the UV background had not yet been switched on; 
ii) HI gets ionized soon after the photoheating was triggered at $z \sim 10$; 
iii) HI mass recovers with the further gas accretion and cooling; 
iv) HI mass gradually decreases along with the slow halo growth at late times. 

Halo 26 is a special case that moved into a dense environment between $z \sim 2-5$, and finally went through a major merger at $z=1.82$, as illustrated in the left inset panel. 
The gas inside was heated and stripped during this process, losing almost all HI. 
Therefore, no star formation occurs in this halo even though it reaches the corresponding mass threshold, $M_\mathrm{ST,\,BLF20}$, consistent with the general picture of environmental quenching that has been reported both theoretically and observationally \citep[e.g.,][]{McCarthy2008, Benitez-Llambay2013, Benavides2025, Zhu2025}. 
After this merger, the halo gradually re-cools its gas inside, with its HI mass increasing by a factor of a thousand, despite experiencing minimal gas accretion. 
In comparison, Halo 34 and Halo 35 underwent gas loss after $z \sim 4$. While the loss in Halo 35 can be attributed to ram pressure stripping in a dense environment, Halo 34 has consistently resided in an isolated environment. 
The reason might be the heating from stellar feedback, and the halos’ weakened ability to cool gas as their mass approached $M_\mathrm{ACH,\,Chan24}$. 

In the middle inset panel, we show Halo 34’s HI density map at 
$z=4.66$, when HI mass peaks at $M_\mathrm{HI}=4.04\times10^7\,M_\odot$. 
We find the high projected HI density region ($n_\mathrm{H} > 10^{19}\,\mathrm{cm}^{-2}$) extends to $\sim$ 3 kpc, exceeding half of the virial radius. 
This leads to a peak star formation rate of $\sim\,0.002\,M_\odot\,{yr}^{-1}$, much higher than the other two cases, corresponding to the exact period the halo mass exceeds the predicted halo mass threshold for star formation $M_\mathrm{ST,\,BLF20}$. 
After $z=4.66$, the threshold mass catches up with the halo mass, and the star formation activity recedes but not completely quenched, possibly due to the gas metal enrichment associated with star formation. 

The right inset panel shows an interesting moment of Halo 35 also at $z=4.66$, when it travelled across a filament-like structure with a nearby massive halo: the HI map becomes temporarily aspherical, particularly in its outskirts at a projected distance of $\sim 3$ kpc, coinciding with the lopsided bulge in Cloud-9's image \citep{Benitez-Llambay2024}. 
After this encounter, the halo moved to a more isolated region and recovered to a more spherical shape, until a similar transition occurred again at $z=0.68$ (not shown). 
Compared to halos with similar mass in the literature, this might be caused by ram pressure stripping in the dense environment \citep[e.g.,][]{Benitez-Llambay2013, Benitez-Llambay2024, Herzog2023} or supernova feedback \citep[][]{Rey2022}, which may require further simulation runs with more snapshots to distinguish between these possibilities in our future work. 

With these three cases, we find halo mass is a crucial but not the only factor in creating and affecting HIDEs: 

i) A moderate mass accretion history that lies between the mass threshold for atomic cooling and star formation, ensures a sufficient level of hydrogen cooling and a minimal stellar mass; thus, the diversity in mass assembly history introduces scatter in their properties. 

ii) Star formation further introduces competing effects: stellar feedback which heats the central gas and suppresses subsequent star formation, and metal enrichment and diffusion which facilitate gas cooling. 

iii) Major mergers, and interactions with passing massive halos or dense environments, can also strongly disturb the internal gas, including heating and stripping via ram pressure, and subsequently reshaping the HI content. 

iv) As each of the above effects sets a timescale for halos to respond, the internal gas may not always be in hydrostatic equilibrium but instead in a transient, evolving state. 

We note that there are other mechanisms which may attribute to additional diversity but are not considered in this work, for example, gas low-temperature cooling, which may trigger hydrodynamical instability and collapse into molecular cloud to form stars, and local UV strength, which relates to different photoheating history that galaxies exposed to and introduces difference in phase diagram, and subsequent star formation history and HI density profiles. These effects are left for our follow-up studies to investigate.

\section{Conclusion} 
\label{sec:conclusion}

In this work, we study HI-rich `dark' galaxies (HIDEs) in the \textsc{hestia} and Auriga simulations, as analogues of Cloud-9 and Leo T. 
halos in these simulations generally contain more gas and stars, and the gas mass vs. halo mass relation inside completely starless halos shows a continuum, rather than the two distinct populations seen in the Apostle simulations (\citetalias{Benitez-Llambay2017}): RELHICs (REionization-Limited HI Clouds, which follow a tight relation between $M_{200}$ and $M_\mathrm{gas}$) and COSWEBs (COSmic Web Stripped systems). 
We select galaxies that are HI-rich, also faint enough to be possibly missed in optical surveys: ($M_\mathrm{HI} > 10^5 \, M_\odot$,  $g$-band magnitude $M_g>-10$), and find 49 analogues in three \textsc{hesita} runs, 34 in six Auriga L3 runs, and 6 in one Auriga L2 run, only one of which is completely starless (Fig. \ref{fig:fig1}). 

The general properties of HIDEs converge well between \textsc{hestia} and Auriga L3 runs despite the mass resolution contrast of $\sim$10 times. 
We find HIDEs have $M_{200} \sim 10^{9.5}\,M_\odot$, $M_\mathrm{gas} \sim 10^{7.4}\,M_\odot$, $M_\mathrm{HI} \sim 10^{6.5}\,M_\odot$, overlapping with the mass range of RELHICs, but they also contain stars of $\sim 10^{5.6}\, M_\odot$. 
They are generally metal-poor in gas ($Z_\mathrm{gas} \sim 10^{-1.4} Z_\odot$), not or barely star-forming, and host old stellar populations formed on average $\sim$~11 Gyr ago (Fig. \ref{fig:fig2}). 
A few of them reside in dense environments as close as $\sim$300 kpc from the Milky Way, coinciding with the distance of Leo T ($\sim$~409 kpc). 

The gas phase diagrams of HIDEs deviate from those of RELHICs, owing to differences in photoheating (self-shielding) prescriptions, stellar feedback, metal enrichment, and environmental diversity (Fig. \ref{fig:fig4}). 
Consequently, these processes introduce diversity of gas and HI properties (such as their density profiles, Fig. \ref{fig:fig5}) which cannot be fully explained by halo mass and concentration. 

To study the origin of HIDEs, we trace their histories and find that their halo masses almost always lie between the thresholds for atomic cooling and star formation.
This regime ensures sufficient radiative cooling, allows them to retain gas, but does not trigger significant star formation.
Once the halo mass approaches or exceeds the star formation threshold, stars form, accompanied by stellar feedback and metal enrichment, which further enhance the diversity.
Interactions with massive halos and dense environments can also heat or distort the HI content, creating aspherical HI maps, as observed for Cloud-9 (Fig. \ref{fig:fig6}).
We conclude that both halo mass history and environmental history are crucial in shaping HIDEs and their diversity.

\begin{acknowledgments}
We thank Alejandro Ben{\'\i}tez-Llambay for sharing the \textsc{Py-SPHViewer} code, and thank Robert J. J. Grand, Mark Krumholz, Jinning Liang, Pablo Renard Guiral, Till Sawala and Jing Wang for useful discussions. 
We acknowledge the support from the National Natural Science Foundation of China
(No. 12473007, 12473015), Beijing Natural Science Foundation (QY23018), and the China Manned Space Program with grant No. CMS-CSST-2025-A03 and No. CMS-CSST-2025-A09. 
This work used the DiRAC@Durham facility managed by the Institute for Computational Cosmology on behalf of the STFC DiRAC HPC Facility (www.dirac.ac.uk). The equipment was funded by BEIS capital funding via STFC capital grants ST/K00042X/1, ST/P002293/1, ST/R002371/1 and ST/S002502/1, Durham University and STFC operations grant ST/R000832/1. DiRAC is part of the UK National e-Infrastructure. 

\end{acknowledgments}


{\it Software:} \textsc{matplotlib} \citep{Hunter2007}, \textsc{numpy} \citep{Harris2020}, \textsc{Py-SPHViewer} \citep{alejandro_benitez_llambay_2015_21703}, \textsc{scipy} \citep{Virtanen2020}.

\bibliography{main}{}
\bibliographystyle{aasjournal}

\end{document}